\documentclass[runningheads]{llncs}
\usepackage{fontawesome}
\usepackage{tabularx}
\usepackage{rotating}
\usepackage{enumitem}
\usepackage{graphicx}
\usepackage{array}
\usepackage{calc} 
\usepackage{microtype}
\usepackage{caption}

\usepackage{xcolor} 
\usepackage[colorlinks=true, linkcolor=blue, urlcolor=blue, citecolor=blue]{hyperref}

\begin{document}
\setlist[itemize]{noitemsep, topsep=0pt, leftmargin=*}
\newlength{\smallcolwidth}
\setlength{\smallcolwidth}{0.25\textwidth} 
\newlength{\bigcolwidth}
\setlength{\bigcolwidth}{0.7\textwidth} 
\captionsetup{belowskip=-10pt}

\title{Exploring a Behavioral Model of “Positive Friction” in Human-AI Interaction}
\titlerunning{Exploring a Behavioral Model of “Positive Friction” in HAI}
\author{Zeya Chen\orcidID{0009-0005-8848-7750} \and
Ruth Schmidt\orcidID{0000-0002-9390-8469}}
\authorrunning{Z. Chen and R. Schmidt}
%
\institute{Institute of Design (ID), Illinois Institute of Technology, Chicago, IL 60616, USA
\email{zchen103@id.iit.edu}}
\maketitle              
\begin{abstract}
Designing seamless, frictionless user experiences has long been a dominant trend in both applied behavioral science and artificial intelligence (AI), in which the goal of making desirable actions easy and efficient informs efforts to minimize friction in user experiences. However, in some settings, friction can be genuinely beneficial, such as the insertion of deliberate delays to increase reflection, preventing individuals from resorting to automatic or biased behaviors, and enhancing opportunities for unexpected discoveries. More recently, the popularization and availability of AI on a widespread scale has only increased the need to examine how friction can help or hinder users of AI; it also suggests a need to consider how positive friction can benefit AI practitioners, both during development processes (e.g., working with diverse teams) and to inform how AI is designed into offerings. This paper first proposes a ‘positive friction’ model that can help characterize how friction is currently beneficial in user and developer experiences with AI, diagnose the potential need for friction where it may not yet exist in these contexts, and inform how positive friction can be used to generate solutions, especially as advances in AI continue to be progress and new opportunities emerge. It then explores this model in the context of AI users and developers by proposing the value of taking a hybrid ‘AI+human’ lens, and concludes by suggesting questions for further exploration.

\keywords{Friction \and Behavioral design \and Artificial intelligence \and Behavioral Science \and  Human-AI interaction.}
\end{abstract}
\section{Introduction}
There is often an underlying presumption when designing for behavioral change that friction is bad, reinforced by advice to ‘make it easy’ or otherwise reduce behavioral barriers as a relatively low-cost, minimally intrusive way to help people act in their own best interests\cite{BIT_2014}. However, seeing friction as uniformly undesirable is too simple. While many applications of ‘make it easy’ remain useful and valid, too much ease can escalate issues of self-control or impulsive behavior\cite{Duckworth_Milkman_Laibson_2018}. Further still, judicious and intentional addition of friction can productively slow down behaviors in a variety of ways, whether by interrupting autopilot behaviors, “cooling off” hot-state decisions through tactics such as waiting periods, or building confidence or rapport by prioritizing interactions over automated efficiency\cite{Schwartz_2022}. 

Purely from a behavioral perspective, therefore, systematically examining how and when friction can yield positive benefits can be useful in situations when speed or convenience tend to dominate problem-solving attention . This is especially pertinent for generative AI and machine learning (ML) platforms and tools, whose recent proliferation has surfaced a new set of tensions between efficiency and reflection both for users and developers of these offerings. This paper first positions the use of friction in a broader behavioral context and presents a behavioral framework that breaks positive friction into various forms. It then applies this framework to real-world examples to illustrate how friction can play a role in AI-informed systems and human-AI interactions (HAI) through characterization, diagnostic, and generative modes, and concludes by proposing future directions.

\section{Behavioral Design in Human-AI Interaction}
The field of behavioral science is often underrepresented in HAI studies, particularly compared to the focus on persuasive technologies in Human-Computer Interaction (HCI) research\cite{Faiola_2007,lee_kiesler_forlizzi_2011b}. Where persuasive technologies primarily aim to alter or influence human behavior through technological and design principles\cite{harris_islam_qadir_khan_2017}, behavioral science encompasses a broader spectrum of disciplines that pulls from psychology, sociology, and economics to understand human decision-making and behavior, and use these insights to design intervention environments\cite{wiese_pohlmeyer_hekkert_2020}. This concept of “behavioral design” merges behavioral science with design research to address complex challenges, emphasizing problem framing before solution development to transcend traditional product-centric design and envision broader interaction experiences, choice environments, and cultural contexts\cite{schmidt_2020a}.

However, behavioral research is increasingly seen as supplying a valuable perspective on understanding and designing for the dynamics of AI within broader socio-technical systems in Human-AI Interaction contexts. Where behavioral science historically has concerned itself exclusively with human behavior, contemporary HAI perspectives that emphasize viewing AI as interactive agents rather than mere technological artifacts underscore the importance of applying a behavioral lens to study their behavioral patterns\cite{mills_2022}. Recent HCI scholarly debates regarding AI anthropomorphism also advocate for a balanced behavioral understanding in AI development that bridges ethical responsibility and the synergies between human and non-human intelligence\cite{tan_2023}. This more systemic and behaviorally-informed approach to AI/human hybrid activities positions it as a promising direction for addressing the increasingly complex, adaptive nature of human-AI interactions within structures, challenging researchers to ensure AI is integrated thoughtfully into these systems.

Therefore, this research positions AI as interactive intelligent agents within socio-technical systems, going beyond traditional computing roles to emphasize its dynamic role in reshaping interactions and solving problems, echoing scholars like Mills\cite{mills_2022} and Shneiderman\cite{shneiderman_2022}. By embracing a behavioral design perspective, we aim to ensure AI's ethical integration into society, fostering responsible interactions between humans and AI.

\section{A Model for Positive Friction}
Friction-reducing strategies and ‘nudges’ include examples such as the Save More Tomorrow (SMarT) program, which leverages human tendencies toward effort aversion by requiring a user to deliberately opt out of auto-enrollment and uses default settings to remove the need to explicitly make a choice\cite{thaler_benartzi_2004}. Similarly, studies have found that healthcare behaviors such as getting flu shots or vaccinations can also benefit from reducing behavioral friction in the form of just-in-time targeted information through text messages\cite{cutrona_golden_goff_ogarek_barton_fisher_preusse_sundaresan_garber_mazor_2018}. But making hard things too easy can result in an inability to exercise self-control. Amazon’s unquestionably convenient One-Click feature removes useful pause points from online shopping, which can easily result in impulsive purchases\cite{kim_2021}. Similarly, waiting periods serve the useful purpose of forcing us to reflect, even if briefly, before committing to a path of action\cite{luca_malhotra_poliquin_2017}. At a systems scale, an absence of friction can have significant negative effects when individual actions aggregate into critical mass movements, such as when easy access to credit and mortgage loans resulted in 2009’s real estate market crash and spate of underwater mortgages, in which property owners found themselves saddled with properties that were worth less than the mortgage they owed\cite{Schwartz_2022}. 

However, positive friction is not a one-size-fits-all proposition. Introducing behavioral speed bumps can provide a stalling mechanism for impulsive behavior, but it can also interrupt auto-pilot behaviors, highlight opportunities or information that might otherwise be overlooked, and provide important feedback\cite{luo2023association}. Looking beyond efficiency can also reveal alternative forms of value, as when the comparatively inefficient mode of train travel reframes getting from Point A to Point B into an experiential benefit compared to traveling by air\cite{dyson_sutherland_2021}. 

\begin{figure}
\centering 
\includegraphics[width=0.89\textwidth]{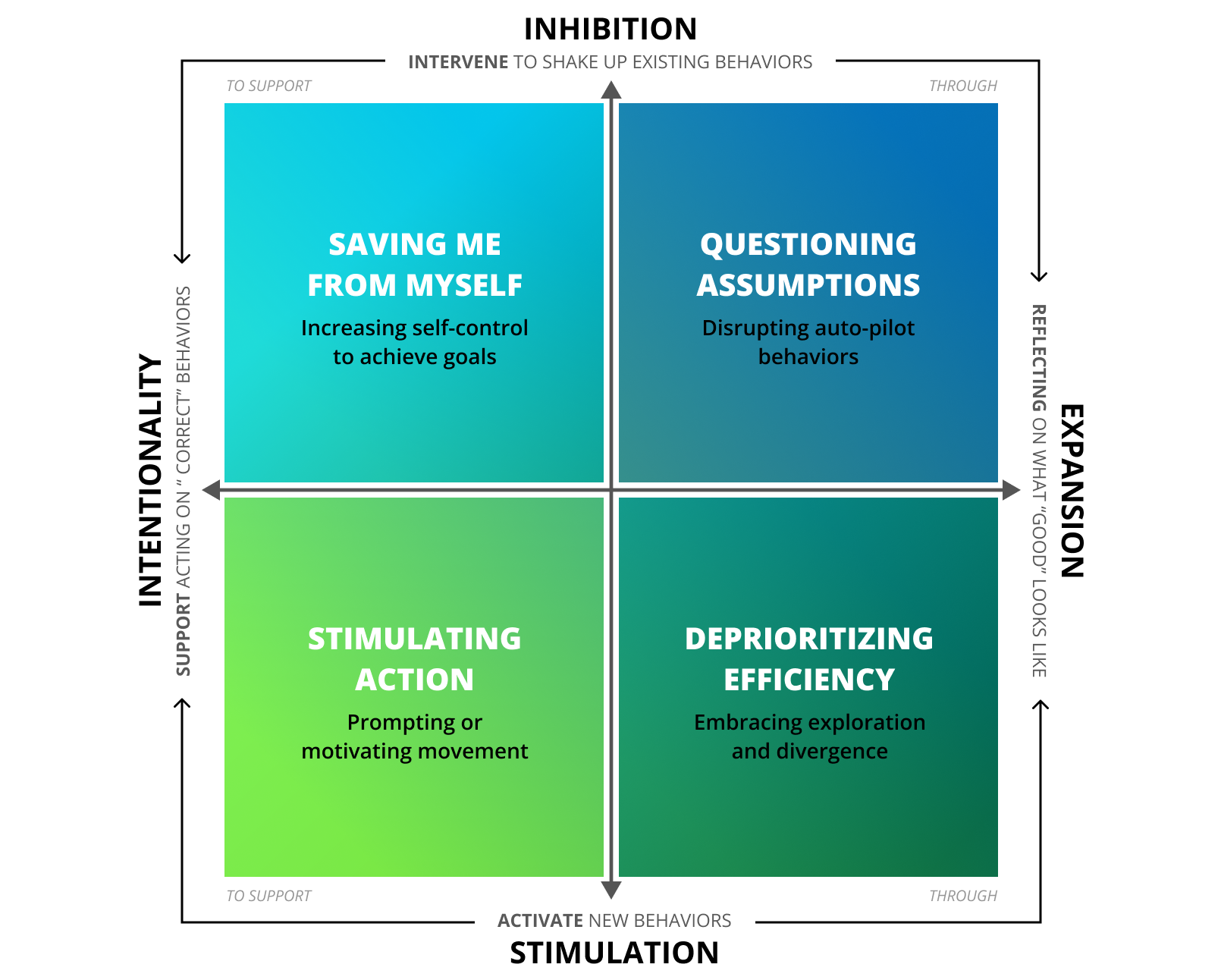} 
\caption{Mapping dimensions of positive friction as activities in service of inhibition vs. stimulation against resultant intentionality vs. expansion of purpose.}
\label{fig1}
\end{figure}

Breaking down the forces behind positive behavioral friction into two opposing axes—inhibiting vs. stimulating actions and goal-based, intentional motivations vs. expanding to see new possibilities—can help differentiate these various forms (Fig.\ref{fig1}). Below, we briefly explain each quadrant and provide examples before exploring how positive friction manifests in a generative AI context.

\subsection{Saving me from myself: Increasing self-control to achieve goals}
Self-control is a well-known behavioral challenge across a wide variety of settings, including healthcare, purchasing, and social interactions, in which individuals struggle to overcome “hot state” temptations if even their more reticent “cold state” selves fully know the downsides of impulsive behaviors\cite{loewenstein2005hot}. As a result, ‘save me from myself’ interventions that intentionally insert logistical, physical, or temporal friction can curtail impulsive tendencies and interrupt undesirable behaviors. These cases of self-control are characterized by clear goals and concrete motivation to change, which can be strategically leveraged by behavioral designers to help individuals avoid temptation. Tactics to help people resist doing things they know they should not may include limiting access to content or engagement, such as reducing product features; imposing timeouts or waiting periods to dampen impulsive actions; adding layers of social accountability or oversight; or introducing self-imposed de-escalation techniques to modulate behavior.

\subsection{Questioning assumptions: Disrupting auto-pilot behaviors}
A second form of positive friction is less concerned with promoting self-control over impulsive behavior than overcoming personal ‘cruise control’ that numbs, rather than heightens, impulses in the form of subconscious or autopilot behaviors. While autopilot behaviors can helpfully reduce cognitive effort, they can lead to tuning out important environmental signals or overlooking opportunities to deviate from the norm even in instances where this may be beneficial. In this case, positive friction may be less oriented toward eliciting a specific preferred outcome, and more toward introducing a greater openness and attention to a full range of options. Behavioral tactics to broaden individuals’ active attention and receptivity to inputs include examples such as modal dialog boxes that prompt reflection before quitting applications,  strategies for ‘inoculating’ individuals against disinformation by pre-empting their exposure to false news\cite{roozenbeek2022psychological} or disfluency techniques that employ intentionally malformed fonts to slow the pace of information processing\cite{benartzi_shlomo_2015}. 

\subsection{Stimulating action: Prompting or motivating movement}
In addition to dampening tendencies toward impulsive or autopilot behaviors, positive friction can also be used to stimulate or encourage new behaviors by providing prompts that energize individuals to take action. This can be especially helpful in cases where adopting good behaviors is challenging due to effort aversion or where an abstract promise of logical gains (e.g., “good health,” “sustainability”) does not supply sufficient motivation. In this case, strategies that employ social norms—such as commitment contracts—can serve as forcing mechanisms to stimulate behaviors by increasing external sources of accountability\cite{Halpern_Asch_Volpp_2012}. Behavioral activation can also occur through increasing the concreteness of both actions and potential benefits; for example, where novel or highly experiential activities can increase engagement by motivating action, focusing on specific stories or individuals, known as the identifiable victim (or actor) effect, can help us overcome the malaise of generic data\cite{jenni_loewenstein_1997,lee_feeley_2016}.

\subsection{Deprioritizing efficiency: Embracing exploration and divergence}
Finally, broad societal tendencies to valorize efficiency often encourage solutions that generate maximal value for minimal effort. However, this can lead to over-indexing on speed and precision over other valuable outcomes and overlook possibilities to create or recognize new forms of value\cite{Schwartz_2022}. Positive friction that disrupts efficiency in productive ways can play an important role in reframing the value of experiences and heightening opportunities to capture these benefits. Behavioral theorists have suggested that amplifying alternative and qualitative aspects of an experience, rather than relying only on quantifiable indicators,  can help surface new forms of value and overcome default tendencies toward efficiency. These can include instances of ‘stealth health’ seen in Pokemon Go, in which the motivation to exercise is subsumed into gameplay\cite{lee_zeng_oh_lee_gao_2021}, or in activities that intentionally cater to alternative values, such as slow lanes in grocery stores that convert transactional activities into social opportunities\cite{Moran_2023}.

\section{Positive Friction and AI}
Despite the widespread application of positive friction in digital environments to tackle specific behavioral challenges—e.g., “save me from myself” interventions in the case of self-control or impulsive behavior—none of these forms of friction are correlated with or constrained by delivery mechanism; in other words, positive friction is technology-agnostic. However, the specific attributes and complexity of AI technologies’ development, problem-solving settings, and opportunities for application make it an especially important area of inquiry. Having laid out the general behavioral positive friction model above, below we suggest how it can inform the design and deployment of AI-informed technology. 

\subsection{Current Research on Friction in AI}
Many companies now rely on AI as a key component of solutions to craft seamless, personalized experiences and reduce barriers that impede customer journey efficiency\cite{gosline_2022}. For instance, Facebook's recent launch of AI-powered "smart glasses" provides an example of "ultra-low-friction input" that enables users to remain effortlessly connected\cite{facebook_reality_lab_2021}. Retail giants like Amazon, Aldi, and Hudson have also embraced frictionless shopping models, allowing customers to bypass traditional checkout lines\cite{logg2019algorithm}; similarly, in the transportation sector, Hitachi and Genfare have experimented with frictionless mobility with AI-based traffic management to optimize traffic flow, reduce congestion, and expedite travel times\cite{Vennelakanti_2021b,kaled_2019}. 

While the trend of "frictionless" AI and machine learning solutions brings certain advantages, it also introduces significant risks, related to privacy concerns, potential amplification of algorithmic biases, intellectual property issues, and ethical challenges. Tendencies such as algorithm appreciation bias, which reveals a preference for algorithmic advice over human input in certain situations, regardless of the transparency of the algorithm\cite{logg2019algorithm,poursabzi2021manipulating} can lead to overreliance on AI, and underscore the potential pitfalls of excessive appreciation of AI advice in the absence of critical safeguards. These tendencies may be particularly important to consider in light of findings from fields concerned with human cognition that integrate neural and symbolic methods to distinguish between quick, instinctive 'System 1' human processing of information and the slower, more thoughtful 'System 2'\cite{kahneman2011thinking}. While current trends in AI development show a preference for the rapid responses characteristic of 'System 1', favoring technologies that offer immediate, effortless solutions, recent proposals have reintroduced interest in specific types of friction to encourage reflection, conflict, and care in AI systems, essentially integrating slower and more reflective 'System 2'-oriented thinking\cite{Rakova_2023}.

Research into algorithm appreciation, cognitive ease, and strain has informed investigations into the intentional use and design of positive friction across various AI-informed domains and challenges. These include examining computational friction on XAI’s interface to calibrate users' trust in the AI system\cite{naiseh2021nudging}; the use of friction as a cognitive tool to foster the civility of online discourse against false algorithmic information\cite{kozyreva2020citizens}; how friction can inform ‘micro-moments’ of human-computer interaction in impulsive and reflective behavioral perspectives\cite{cox2016design}; and the value of positive friction by Contestational Design for civic engagement\cite{korn2015creating}. Expanding on these experiments and findings from the perspective of strategic behavioral design can help inform and enrich the design of future AI-empowered products and service systems.

\subsection{Exploring the Beneficiaries of Positive Friction in AI}
While designers and developers can build on a long tradition of removing and employing friction in products and services, the widespread and embedded nature of AI in consumer offerings has increased both the need and the stakes to consider the role—and intentional incorporation—of friction in AI solutions. Just as instances of positive friction designed into offerings can target end-user or consumer behaviors to encourage self-regulatory behaviors and cultivate beneficial habits\cite{piras_2023}, AI practitioners can also benefit from behavioral speed bumps during the design and development of AI-enabled solutions by productively slowing down or interrupting design processes or inserting opportunities to identify and rectify biases that algorithms may perpetuate (Table\ref{tab:label1}). However, the specific application of positive friction for both users and developers of AI may differ, depending on the nature of the situation or the intended goal.

\begin{sidewaystable}
\centering
\small
\caption{Illustrative examples of positive friction in AI end-user and practitioner contexts }
\label{tab:label1}
\begin{tabularx}{\textwidth}{ | >{\raggedright\arraybackslash}p{\smallcolwidth} | >{\raggedright\arraybackslash}p{\bigcolwidth} | >{\raggedright\arraybackslash}X | }
\hline
 & \textbf{Positive Friction for AI Users} & \textbf{Positive Friction for AI Practitioners} \\
\hline
\textbf{Saving me from myself:} Increasing self-control to achieve goals & 
\begin{itemize}
  \item ChatGPT’s cautionary message under its input box encourages users to verify information, reducing over-reliance on AI\cite{openai_2022}.
  \item X (formerly Twitter)’s “Read before you post” feature aims to prevent impulsive sharing of misinformation or harmful content\cite{TwitterSupport2020}.
\end{itemize} & 
\begin{itemize}
  \item Adobe’s diverse design team acts as a safeguard, testing Adobe Firefly’s outputs to prevent stereotypical imagery and enhance data diversity\cite{bardlavens_2023}.
\end{itemize} \\
\hline
\textbf{Questioning assumptions:} Disrupting auto-pilot behaviors & 
\begin{itemize}
  \item Grammarly’s feature that highlights gender pronouns in AI-assisted writing helps prevent unconscious misuse of “he/she/they”\cite{yavnyi_2020}.
  \item Studies show that news recommendation AI with ‘opposing views’ features can break echo chambers by suggesting diverse viewpoints\cite{heitz2022benefits}.
\end{itemize} & 
\begin{itemize}
  \item X’s algorithmic bias bounty challenges invite public scrutiny to uncover hidden biases in machine behavior\cite{chowdhury_williams_2021}.
\end{itemize} \\
\hline
\textbf{Stimulating action:} Prompting or motivating movement & 
\begin{itemize}
  \item Tesla’s Autosteer mode requires drivers to hold the steering wheel, enhancing attentiveness and safe driving\cite{TeslaModel3Manual2023}.
  \item Google’s Jigsaw Perspective API prompts users to reconsider potentially harmful comments before posting \cite{PerspectiveAPI2023}.
\end{itemize} & 
\begin{itemize}
  \item Microsoft’s Fairlearn\cite{bird2020fairlearn} and IBM’s AI Fairness 360\cite{bellamy2018ai} are toolkits that help mitigate bias in machine learning models, promoting responsible AI development.
\end{itemize} \\
\hline
\textbf{Deprioritizing efficiency:} Embracing exploration and divergence & 
\begin{itemize}
  \item Academic authors are increasingly required to clarify AI’s contribution in research \cite{stokel2023chatgpt}, fostering exploration of AI’s potential and ethical use.
  \item The “6-month pause on giant AI experiments” open letter\cite{Future_of_Life_Institute_2023} may delay AI development and allow time for human reflection and reaction.
\end{itemize} & 
\begin{itemize}
  \item Google’s decision to end third-party cookies shifts focus from efficient data tracking to protecting user privacy and responsible AI marketing strategies\cite{Chavez_2023}. 
\end{itemize} \\
\hline

\end{tabularx}
\end{sidewaystable}

In addition, different stakeholders may have diverse interactions, values, and contexts of use that require different strategies and considerations for when, why, and for whom positive friction is used. This parallels findings in the recent AI-related socio-technical research highlighting the significance of clearly defining and considering not only multiple stakeholder groups, including those who contribute to, develop, deploy, and consume or are otherwise affected by AI,  but by development stages—such as data collection, training, model evaluation and analysis, and system deployment—when developing AI regulatory oversight\cite{Mitchell_2023}. 

Considering positive friction at multiple levels can yield improved products and experiences, as illustrated through two examples of ‘X’ (formerly known as Twitter). The “Read before you post” feature\cite{TwitterSupport2020} applies positive friction as a social media safeguard to deter impulsive sharing of misinformation or harmful content. While this feature effectively shields some participants from harmful or deceptive information, its impact is limited to resharing and replying by users. As a result, the use of friction is reactive rather than preventive given that it does not extend to the creators of harmful content or the biased algorithms that promote such content. 

In contrast, ‘X’s' Algorithmic Bias Bounty Challenge\cite{chowdhury_williams_2021} represents an example of employing positive friction more proactively by managing misinformation closer to its origination by targeting AI developers\cite{chowdhury_williams_2021}. In this case, the Challenge founders shared ‘X’s' saliency algorithm (also known as Twiitter’s image cropping algorithm), making their code available for participants in order to reproduce and examine bias issues in the algorithms at DEFCON, a prominent hacker conference\cite{chowdhury_williams_2021}. The results of this challenge showed that participating hackers succeeded in revealing several hidden biases inside ‘X’s’ AI systems for future debugs, including but not limited to encoded stereotypical beauty standards, algorithmic preferences for certain race features, and linguistic biases\cite{yee_peradejordi_2021}. 

Unlike the purely user-focused “Read before you post” feature, the Bounty Challenge introduced friction within the platform itself; this not only prompted AI developers to reassess and refine machine functions, but also proactively addressed AI algorithmic issues in advance rather than designing positive friction features that users would need to use reactively. While both examples share the objective of fostering a less biased and more inclusive online environment, their differences illustrate how effective use of positive friction in an AI context not only can affect different stakeholders in varying contexts but highlight that different forms of positive friction serve distinct roles. Identifying relevant stakeholder groups can help designers better understand the benefits, harms, and risks presented by different AI applications, thus guiding the development of more effective and precise positive friction strategies and avoiding a one-size-fits-all approach that is inadequate for complex issues surfaced by advances in AI.

\subsection{Collaborative Dynamics of Positive Friction in AI+Human Hybrids}
\begin{figure}
\centering
\includegraphics[width=0.85\textwidth]{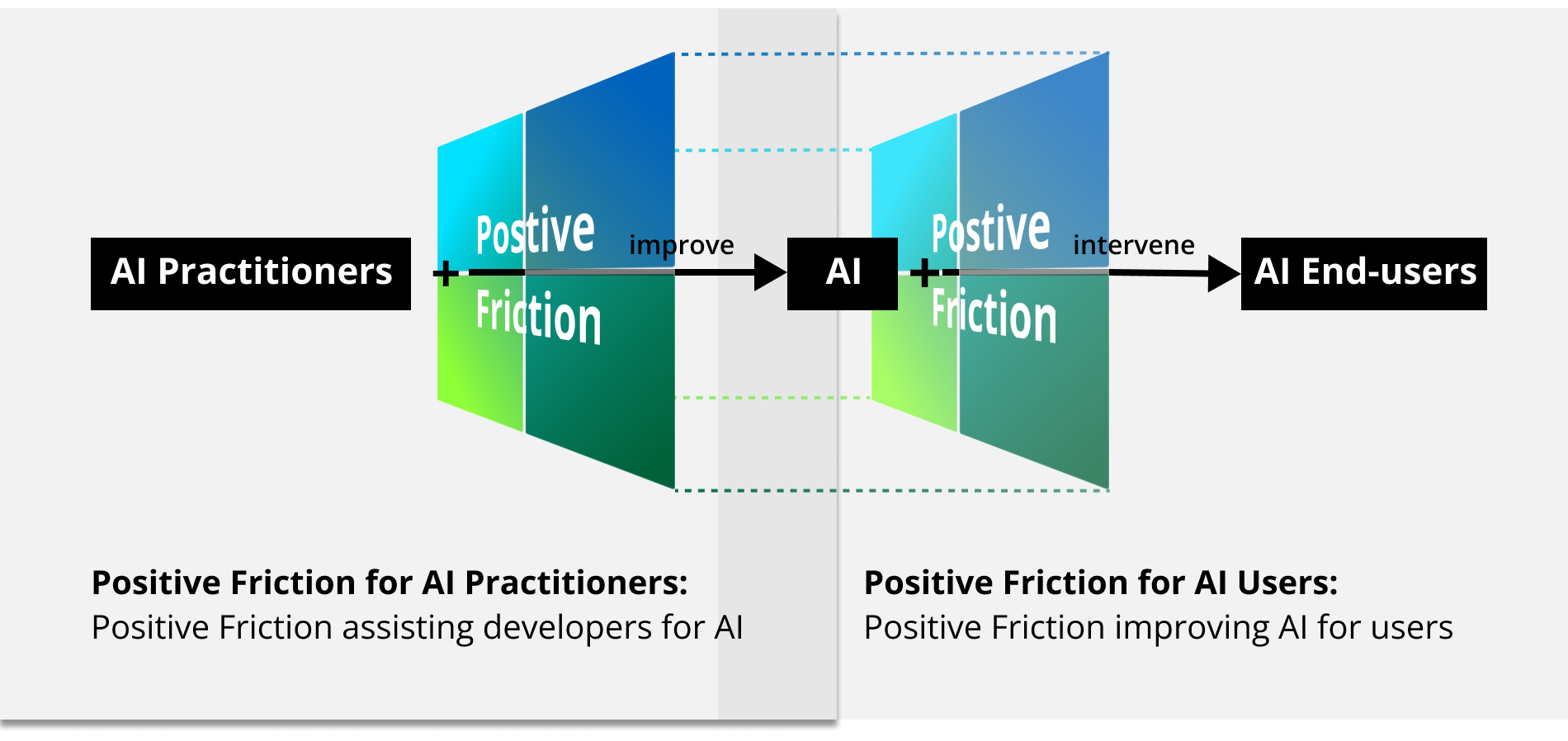}
\caption{Grammarly’s mutual supportive “practitioner-AI-user” relationship with Positive Friction Model.} \label{fig2}
\end{figure}

Recognizing the stakeholders involved also underscores the interconnectedness between AI technology, AI practitioners and users. Frequently, efforts to integrate AI into society foster a “competitive, human-machine opposition mindset,” either through technical approaches to improve the algorithms themselves outside of human oversight or through governance and oversight that create regulations around algorithms, both of which isolate AI technology from human operators (AI practitioners), and collaborators (AI users)\cite{guszcza2022hybrid}. By introducing behavioral speed bumps into AI development, however, developers can more easily surface hidden issues in algorithms that can subsequently enhance user experiences and promote healthier human-AI interaction behaviors. This positions positive friction less as a choice between “assisting developers for AI" or “improving AI for users" or reinforcing a divide between AI developers and end-users, than fostering a collaborative “practitioner-AI-user” dynamic flow through responsible integration of technology and human experience (Fig.\ref{fig2}). In short, incorporating positive friction into both the design and deployment of AI can reframe the antagonistic dynamic between human agents and algorithms into a  collaborative one,  with the potential for human-machine hybrid intelligence. 

Grammarly’s recent research on gender-inclusive grammatical error correction in natural language processing (NLP) systems\cite{lund2023gender} presents an example of this kind of multi-faceted, multi-stakeholder mutual support problem-solving. For AI users, Grammarly’s AI-driven spell-checker and autocorrection function serves as positive friction through pop-up messages questioning personal pronoun usage ('he', 'she', or 'they') or suggestions to replace 'mankind' with 'humankind,' which disrupt writers’ unconscious typing behaviors but encourage them to reconsider pronoun biases, or prompt reflections on gender-neutral language. For AI practitioners, flagging corrections' biases made by the Autocorrection represents positive friction in the form of a trigger during the evaluation of AI, as when Autocorrections suggest changing “Charlie’s earnrings” into “his earnings” or “her earrings”\cite{yavnyi_2020}. Although time-consuming, this use of positive friction to conduct internal checks and 'mine clearance' is not only vital for debugging and improving the AI systems, but also reinforces organizational values in support of LGBTQIA+ inclusivity and a diverse workforce\cite{yavnyi_2020,grammarly_2023}. Grammarly’s commitment to positive friction not just as a device to improve technology but to bolster social and cultural values can be seen in their engineering team’s Counterfactual Data Augmentation(CDA)-supported model for internal Grammatical Error Correction(GEC) systems testing and training, and its subsequent open-source release to the public\cite{lund2023gender}. This initiative not only enhances self-control to mitigate gender bias in Grammarly’s AI systems, but also encourages the inclusive development of linguistic AI on a broader scale. 

The increased prevalence and use of AI requires considering how a diverse range of stakeholders and situations can benefit from positive friction across AI development and end-user engagement. Recognizing that AI/human hybrid behavior as a complex multi-stakeholder unit underscores that identifying the need for positive friction, selecting strategies to address it, and incorporating it into AI, developer, and user dynamics is not a 'one-size-fits-all' solution. The case of Grammarly, for example, illustrates how different forms of positive friction, sourced from various quadrants in the model, can benefit different stakeholders to improve the offering as a whole. In addition, given the wide range of stakeholders, community values, and application environments, a singular design approach to positive friction may resonate differently across various groups, suggesting that AI design requires a full understanding of relevant contexts before implementation, and further that designing friction solely with specific groups like AI users or practitioners in mind can inadvertently overlook the needs and responsibilities of still other stakeholders. This suggests not only the value of exploring the model’s dynamic relationships in combination rather than as four individual strategies or a singular audience, but also the importance of recognizing how different forms of positive friction might be naturally intertwined. However, more is not always better, and the effective use of positive friction does not require using strategies from all four quadrants. Instead, the design and development of AI requires a rich understanding of context, the nature of potential challenges that might benefit from positive friction, and a sense of which strategies are best suited to the task.

\subsection{Positive Friction Model as Lens}
Seeing positive friction in an AI context as a complex design challenge that requires solving for multiple people and problems simultaneously suggests that employing the Positive Friction Model can take three different forms: 1) as a characterizing lens to systematically identify the nature of existing friction, 2) as a diagnostic lens to analyze how situations within complex challenges that would benefit from positive friction, and 3) as a generative lens to craft targeted positive friction that effectively address these challenges.

\subsubsection{Characterizing Lens:} 
When used to characterize existing forms of friction, the Positive Friction Model can help designers better understand where strategies are already in use, who is benefitting from them, and how they might be analyzed to see how they are creating impact. Below, we examine three cases—the Future of Life Institute (FLI)’s Open Letter proposing a pause on AI system training\cite{Future_of_Life_Institute_2023}, Tesla’s auto-driving features\cite{TeslaModel3Manual2023,tesla_2021}, and Grammarly, as introduced above—by capturing and characterizing positive friction features employed in each.

On March 22nd, 2023, FLI published an Open Letter calling for a six-month pause on “the training of AI systems more powerful than GPT-4,” warning of “an out-of-control race to develop and deploy ever more powerful digital minds that no one—not even their creator—can understand, predict, or reliably control”, and asking the world to bask in a “long AI summer, not rush unprepared into a fall”\cite{Future_of_Life_Institute_2023}. The short but highly publicized statement garnered over 30,000 signatures in a few short months\cite{Future_of_Life_Institute_2023}. Characterizing this instance according to the Positive Friction Model provides several immediate insights. First, it can help us break down which various friction-based strategies are in use;  for example, the letter simultaneously applies a self-imposed waiting period in the form of an explicit and lengthy pause, as a kind of self-control mechanism, to slow down AI progression (Save me from myself); introduces skepticism and critical thinking as mechanisms to questioning the tempting convenience of AI (Question assumptions);  uses social norms and concreteness in the form of signatories as prompts for reflection, accompanied by an urgent tone that raises the stakes for AI regulators to take actions (Stimulating actions); and suggests friction in the form diversity by encouraging more inclusive discussions (Deprioritizing efficiency). Second, taking a categorizing lens also allows us to identify not just what strategies are used, but to whom these interventions are directed; in this case,  across three distinct groups, including developers, regulators, and the public\cite{struckman2023they}. Finally, identifying the range of strategies in use and their intended targets allows us to see patterns to identify where and for whom friction is currently being used, and just importantly where it is not.

Tesla’s auto-driving mode offers another example of the value of categorizing existing friction, in the form of Autosteer alerts\cite{TeslaModel3Manual2023} and behavioral-score-driven insurance\cite{tesla_2021}. The use of friction in autosteer mode requires drivers to maintain contact with the steering wheel during auto-steering, triggering visual and auditory alerts if compliance is not detected, using direct feedback to encourage safe driving behavior by stimulating responsiveness and evasive actions as necessary. In contrast, Tesla’s insurance scheme bases premiums on real-time driving safety scores rather than conventional factors like gender and age. This more subtle form of friction encourages drivers to avoid risky behaviors like hard braking, aggressive turning, or phoning while driving, due to the knowledge that overly aggressive behaviors risk incurring penalties in the form of higher insurance rates. While these instances of friction differ in their design—the former explicitly prevents drivers from engaging in dangerous activities where the latter employs implicit punishment for ‘bad’ behaviors, both target AI users and both share the same underlying purpose of increasing drivers’ attentiveness and promoting safer driving behaviors during auto-driving. Positioning these instances within the Model also allows us to recognize the additional insight that both strategies appear on the “intentionality” side of the model, which is perhaps not surprising given they are both highly goal-driven to support specific ‘correct’ behaviors, but which also can spur new thinking about how expanding perceptions might be employed to encourage driver safety.

\begin{figure}
\centering
\includegraphics[width=\textwidth]{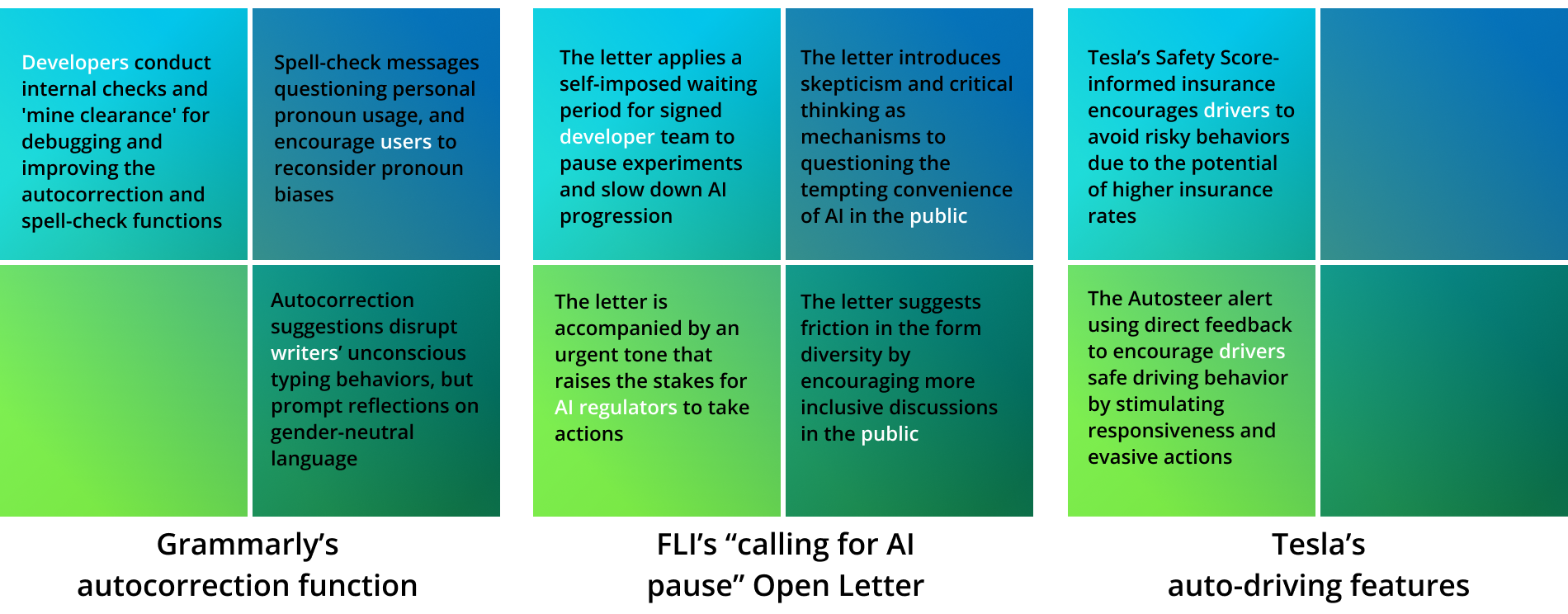}
\caption{Three cases as characterized according to the Positive Friction Model} \label{fig3}
\end{figure}

Combining these instances and insights with the previous example of Grammarly allows us to observe how different cases exhibit unique combinations of positive friction in types, purposes, and target audiences (Fig.\ref{fig3}). It also reinforces that more or more types of friction do not automatically lead to better outcomes if the specific context does not require them or if they are designed to solve different things. For instance, although all four quadrants of positive friction strategies can be seen in the Open Letter case, they address different groups with varied objectives. In contrast, the Tesla case employs only two types of positive friction, but in sharing an intended purpose for a specific user-stakeholder they are able to play a useful reinforcing role. This highlights the important point that merely increasing the quantity of friction is not inherently beneficial and that effective use of friction requires focusing on context-specific designs that enhance the effectiveness of solutions.

\vspace{-15pt}

\subsubsection{Diagnostic Lens:}
The Positive Friction Model can also be used diagnostically to help identify when situations might benefit from positive friction and how to apply it. Where an emphasis on characterization enables us to analyze existing friction, taking a diagnostic lens can help us identify why, when, and where positive friction should be introduced in new problem contexts, or to augment where the existing use of friction may not go far enough.

Consider the FLI's Open Letter; where characterizing current forms of friction identified a wide range of strategies, a diagnostic lens might enable more insight into whether all instances of positive friction were equally effective, and where they might need to be complemented by other interventions. For example,  even signatories who doubted the feasibility of a six-month pause valued the letter's public awareness of AI's rapid advancements\cite{struckman2023they}. This suggests that while the letter supplied a useful form of friction in the form of a conversation starter, additional and more explicit brakes on the system might be necessary to more forcefully stall and reflect on AI development. Similarly, adding friction does not ensure that system participants and stakeholders will respond as expected; this can lead to reactant behaviors that achieve the opposite effect of interventions’ intent, as seen in instances of individuals who perceived the letter as a stimulus accelerating AI development rather than pausing it\cite{leaver2023chatgpt} or concerns about the letter’s potential to increase public AI anxiety or inadvertently promote AI hype for commercial gains \cite{DrTechlash2023}. Similarly, while Tesla’s initial use of positive friction measures to enhance driver attention saw early success, users soon devised ways to trick them, such as using 'wheel weights/knobs' or even oranges to mimic hand pressure on the steering wheel to such an extent that 'wheel weights' were still among the top sellers in Amazon's "automotive steering wheels" category in 2023\cite{siddiqui_2023}. These unintended consequences highlight the importance of testing friction-based behavioral interventions in a rapidly evolving AI landscape.

In these cases, using the Positive Friction Model diagnostically helps us identify new situations and appropriate ways in which to employ positive friction to improve AI design. In the Open Letter case, for example, a diagnostic lens can signal the need to adopt more inhibitory strategies; in Tesla’s scenario, developers might redesign how friction is built into offerings, to avoid situations where the temptation for some users to exhibit control over driving is strong enough that they deliberately game the system. But a diagnostic lens can only do so much, and in some cases the need for positive friction may not reveal itself until after negative consequences have already occurred. For example, despite ChatGPT's cautionary message ("ChatGPT can make mistakes. Consider checking important information")\cite{openai_2022}, an insufficient degree of friction combined with plausible-looking results and the temptation of efficiency have led individuals to believe AI’s “hallucinated” references and citations\cite{wise_2023}. Tesla also was forced to institute recalls when auto-drive features flouted local driving laws, such as inappropriate lane changes or traveling through yellow lights before the driver had the opportunity to intervene\cite{linkov_2023}. These cases reinforce the need to not just solve problems as they occur but learn how to identify where adaptations to friction can fill gaps in current preventative measures and where new approaches—such as regulatory oversight—may be necessary to address new and emerging situations more proactively.

\subsubsection{Generative Lens:}
Finally, the Positive Friction Model can be used generatively to help create structured, multifaceted interventions that target diverse stakeholders both effectively and ethically.  While there is no one-size-fits-all approach to generating positive friction solutions, the specific characteristics of AI offerings and capabilities suggest several key considerations:

\textit{Expanding the Scope of Stakeholder Analysis} - While focusing on AI users and practitioners will continue to be at the core of any positive friction strategy to address AI-related behaviors, exploring a broader array of relevant stakeholders can surface other important challenges that might otherwise go overlooked. In the Tesla scenario, for example,  addressing the ‘wheel weight\cite{siddiqui_2023}’ workaround that overcame initial driver-based-friction interventions might require new approaches that encourage AI developers to decrease machine intelligence and encourage manual control in certain situations to buy time for AI upgrades. Even more dramatically, new positive friction interventions could extend beyond the immediate AI+human (i.e., vehicle+driver) hybrid unit to focus on ‘wheel weight’ manufacturers and retailers by imposing stricter regulations or penalties for selling harmful devices, or on traffic police to incentivize noting the use of wheel weights at traffic stops. 

\textit{Embracing Expansive, Reflective Friction} - Beyond goal-driven interventions, there's value in embracing expansive, reflective friction. For example, Tesla's Autosteer alerts and behavioral-scored insurance may not achieve their intended goals in part due to limitations of interventions solely focused on modifying behavior without enhancing users’ willingness to comply. This may suggest the value in exploring additional forms of friction represented in the model, less as a quest for comprehensiveness or more-is-better attempt to force-fit additional forms than an effort to more holistically gauge the constellation of challenges that various stakeholders may represent. For example, James Bridle’s ‘Autonomous Trap 001’ project\cite{Bridle_2017} cleverly used a salt circle on the ground to deceive an autonomous vehicle’s vision system; this not only sparked debates on AV safety and limitations, but also challenged the overconfidence and overreliance of AV users. 

\textit{Integrating Hard and Soft Friction} - While the Positive Friction Model introduces four specific types of positive friction strategies, it is less specific about the material or form factor that friction might take in practice. Similar to system design’s characterization of hard (e.g., physical, technological) and soft (e.g., cultural, social) system features, instances of friction can also be positioned as hard or soft\cite{nold2021towards}. Hard applications of positive friction usually influence behavior more directly either through physical barriers or technological infrastructures, often employing more top-down or paternalistic strategies leading to specific actions. Examples of this may include Tesla's Autosteer alerts\cite{TeslaModel3Manual2023}, ‘X’s’ “read before you post” feature\cite{TwitterSupport2020}, governance policies, and fine punishments, all of which are structurally embedded in system infrastructures. Conversely, soft friction influences behavior more indirectly by affecting mental attitudes or social norms to encourage user actions. These might include Tesla’s behavioral-scored insurance\cite{tesla_2021}, FLI's celebrity-endorsed Open Letter\cite{Future_of_Life_Institute_2023}, and Bridle's AV trap project\cite{Bridle_2017}, which apply principles like ‘mental avoidance of harm’, ‘social norms’, and behavioral ‘nudges.’ Integrating both hard and soft friction can enhance the overall impact of positive friction. 

\subsection{Implications and Questions for Further Exploration}
While integrating behavioral science and design research approaches like the Positive Friction Model more deeply into HAI practice may enhance HCI approaches to socio-technical studies, merging disciplines faces challenges of methodological and definitional alignment. In addition, advances in AI’s capabilities and use continue to evolve nearly on a daily basis. On the one hand, this research heightened the potential value and urgency of employing a behavioral design lens; on the other, key questions about situations of use and potential applications of behavioral tools, such as positive friction, have yet to be defined or even understood.  However, despite this uncertainty, there are several potential research questions and further exploratory directions for positive friction in AI worth considering even at this early stage.

First, the Positive Friction Model should be seen not as a stand-alone tool but as a complementary approach that can be used in conjunction with established design frameworks (e.g., journey maps, POEMS\cite{kumar2012101}) and behavioral tools (e.g., the COM-B framework\cite{michie2014behaviour}, the Choice Triad\cite{schmidt2022choice}), combining familiar methods in novel ways that can potentially reveal insights not previously considered. Employing mixed methods can balance the reliance on a single discipline, merging data-driven confidence with openness to a broader range of solutions; for instance, the use of journey maps that demonstrate forms of multi-stakeholder involvement may yield insight into postures and attitudes involved in various AI-related activities, providing an experiential perspective that the Positive Friction Model currently lacks. This also suggests the potential value of exploring how HCI frameworks or tools can integrate with the Positive Friction Model, whether using HCI approaches to enhance the implementation of positive friction from strategic principles to practical experiments or how targeted use of positive friction can benefit HCI solutions. 

In addition, no methodology is entirely neutral\cite{winner2017artifacts}; every designer, framework, and solution inherently carries biases, and even methodologies that are scientific and objective (such as behavioral science), openly participatory (as design strives to be), or data-driven and evidence-based (as HCI studies) are influenced by underlying ideologies that dictate disciplinary norms, what qualifies as evidence, and what constitutes a successful solution. Diversity is itself a form of friction, suggesting that successful collaborations across disciplines can benefit from the insertion of deliberate pause points or stimulations that prompt seeing old ideas in new ways. The Positive Friction Model, therefore, can play a significant role within the design process, in which it can inform and support cross-disciplinary exchange as a dialogic tool that helps transdisciplinary teams to identify tensions or blind spots, foster continual conversation, encourage engagement with diverse viewpoints, and position design as an iterative exchange rather than a final decision-making point. 

Finally, effective use of the Positive Friction Model and positive friction strategies may change over time as AI itself evolves. Just as the widespread use of LLMs such as chatGPT and visual engines such as DALL-E and Midjourney in the Fall of 2023 introduced both fascination (of their ‘automagical’ abilities to conjure convincing and well-structured content and media) and fear (of AI taking over jobs or concerns about cheating in educational settings), new and emergent forms of AI will also arise. In addition,  the distinction between users and practitioners may blur as AI evolves, requiring new or adaptable design strategies to accommodate new use cases that may be difficult to identify in advance. Given that designers are likely to have a front seat as these new technologies are ushered in, designers will also have the ability to influence how AI is seen and adopted by people and more broadly by society\cite{eggink2022tool}. With this ability comes the responsibility of proficiently employing positive friction across the arc of AI development, implementation, and use;  minimizing its potential to create unintended consequences, and updating why, when, and how strategies are used in future contexts or to address future challenges in ethical and equitable ways.

\section{Conclusion}
Despite the fact that friction is traditionally seen as a negative when designing for behavior, positive friction can be used in situations requiring greater self-control, higher reflexivity, and contexts in which efficiency and autopilot behaviors may dominate. The application of positive friction may be increasingly useful in the case of both AI development, in which frictionless processes can too easily increase the likelihood of algorithmic bias or misuse, and for AI users, who may lack the knowledge or ability to adequately question AI outputs. This may be particularly necessary given the recent theory that positions ‘human+AI’ not merely as users and tools, but as hybrid agents that complement and supplement each others’ strengths in a wide range of personal and professional settings\cite{guszcza2022hybrid}.  

However, successful use of positive friction requires the ability to characterize when it is already being used in offerings and products, how to diagnose its effectiveness or identify challenges that would benefit from additional attention, and a generative lens that can integrate positive friction into structures, processes, and offerings as necessary. In the case of AI, this means being both reactive to emergent issues—such as algorithmic biases that cause harm or human behaviors in need of adjusting—as well as an openness to proactively identifying new opportunities where friction may be beneficial, especially as technology continues to advances in as yet undetermined ways. This also requires recognizing that positive friction is not a one-size-fits-all approach, but a systematic way to design for multiple stakeholders, values, and complex systemic contexts of AI, ensuring that AI systems are not just efficient and technologically sound but also mindful of societal values and user welfare.  While AI technology and its use will assuredly continue to evolve,  it is equally likely that positive friction will continue to demonstrate adaptability and relevance in an evolving technological landscape. As AI continues to evolve, we believe the concept of positive friction can help steer this progression toward beneficial and equitable outcomes as well.

%
%
%

%

\end{document}